\documentclass[%
 reprint,
 amsmath,amssymb,
 aps,
]{revtex4-2}

\usepackage{graphicx}
\usepackage{dcolumn}
\usepackage{bm}
\usepackage{amsmath}
\usepackage{comment}
\usepackage{color}

\begin{document}

\preprint{APS/123-QED}

\title{
Sparse modeling approach for quasiclassical theory of superconductivity
}

\author{Yuki Nagai}
\affiliation{CCSE, Japan Atomic Energy Agency, 178-4-4, Wakashiba, Kashiwa, Chiba, 277-0871, Japan}
\affiliation{
Mathematical Science Team, RIKEN Center for Advanced Intelligence Project (AIP), 1-4-1 Nihonbashi, Chuo-ku, Tokyo 103-0027, Japan
}
\email{nagai.yuki@jaea.go.jp}

\author{Hiroshi Shinaoka}
\affiliation{Department of Physics, Saitama University, Saitama 338-8570, Japan}
\affiliation{JST, PRESTO, 4-1-8 Honcho, Kawaguchi, Saitama 332-0012, Japan}

\date{\today}

\begin{abstract}
We propose the sparse modeling approach for quasiclassical theory of superconductivity, which reduces the computational cost of solving the gap equations. 
The recently proposed sparse modeling approach 
is based on the fact that the Green's function has less information than its spectral function and hence is compressible without loss of relevant information. 
With the use of the so-called intermediate representation of the Green's function in the sparse modeling approach, one can solve the gap equation with only 10-100 sampled Matsubara Green's functions, while the conventional quasiclassical theory needs 100-1000 ones.  
We show the efficiency of our method in bulk and vortex states, by self-consistently solving the Eilenberger equations and gap equations. 
We claim that the sparse modeling approach is appropriate in all theoretical methods based on the Matsubara formalism in the quasiclassical theory of superconductivity. 
\end{abstract}

\maketitle


\section{Introduction}
The quasiclassical theory of superconductivity is successful in the weak coupling Bardeen-Cooper-Schrieffer (BCS) model of superconductivity\cite{kopnin2001theory}.
The theoretical framework is based on the fact that the coherence length $\xi$ is sufficiently greater than the Fermi wavelength $1/k_{\rm F}$, i.e., $\xi k_{\rm F} \gg 1$ which expresses a typical scale difference in weak-coupling superconductivity. 
Various kinds of analytical and numerical techniques on the quasiclassical theory have been developed and successfully applied to the studies of a large number of conventional and unconventional superconductors\cite{kopnin2001theory,eilenberger,Schopohl,Hayashiimp,HAYASHI201369,KramerPesch,melnikov2008,miranovic2004,nagai2006,nagai2016,nagaijpsj2006,nagaiprb2012,nagaiprl2008,volovik,nagaijpsj2014}. 

The self-consistent calculation of the gap equation in the quasiclassical theory has been used for studying interesting inhomogeneous phenomena in conventional and unconventional superconductors. 
For example,  a spontaneous symmetry breaking has been studied at at surfaces of $d$-wave superconductor\cite{hakanssonnatphys,wennerdal,holmvall}. 
The theory of the magnetic-field-angle dependence of the heat capacity, which is one of the most powerful tools to study superconducting pairings, has been used in various kinds of vortex lattices in unconventional superconductors\cite{an2010,doi:10.1143/JPSJ.79.094709}. 

On the other hand, it is well known that the quasiclassical theory with full self-consistent calculations, ({\it i.e.} calculation with solving gap-equations, Dyson equations for self-energies, and Maxwell equations) is expensive in  inhomogeneous systems. 
The quasiclassical theory is usually used in a lower temperature region, since there is a numerically less expensive Ginzburg–Landau theory near the critical temperature. 
In the low temperature region, we need to introduce a large number of Matsubara frequencies to self-consistently solve the gap equations. 

Recently, one of the authors and co-workers have proposed a physically motivated compact representation of the imaginary-time and Matsubara Green's function\cite{shinaokaPRB2017,doi:10.7566/JPSJ.89.012001,ir-review,PhysRevB.103.045120,doi:10.7566/JPSJ.88.064004,itou2020}\footnote{For a review, refer to Ref.~\onlinecite{ir-review}.}. 
The formalism is based on the fact that extracting the spectral function from the imaginary-time Green's function is an ill-posed problem. 
This means that the Green's function has less information than the spectral function and hence is compressible without loss of relevant information. 
This approach utilizing the sparseness of imaginary-time quantities is called the \textit{sparse modeling approach}. 
By introducing the compact and efficient representation (which we call intermediate-representation (IR)) of the Green’s functions, 
it has been shown that the conventional uniform Matsubara frequency grid can be replaced by a set of sparse sampling points that can describe the frequency dependence of the IR basis and hence Green’s functions\cite{PhysRevB.101.035144}.
Using the sparse sampling method, we can reconstruct the Matsubara Green’s function with only about 100 points on the frequency grid, and transform efficiently the imaginary time Green’s function to the Matsubara Green’s function and vice versa.
With the use of this sparse sampling method, the {\it ab initio} Migdal-Eliashberg calculation is efficiently solved with the IR basis\cite{PhysRevB.102.134503}.

In this paper, we propose the sparse modeling approach for quasiclassical theory of superconductivity, which reduces the computational cost of solving the gap equation. 
The quasiclassical Green's function, a central object in the quasiclassical theory of superconductivity, is determined by a contour integral of the conventional Green's function. 
We find that the quasiclassical Green's function itself can not be expanded by the IR basis. 
However, in terms of solving the gap equations, we show the way to use the IR basis. 
With the use of the sparse modeling approach, we show that the gap equation can be solved with only 10-100 sampled Matsubara Green's functions. 

This paper is organized as follows. 
In Sec.~\ref{sec:quasi}. we introduce the quasiclassical theory of superconductivity. 
In Sec.~\ref{sec:spm}, we briefly introduce the sparse modeling approach. 
In Sec.~\ref{sec:gap}, we show how to use the sparse modeling approach for solving the gap equations. 
We introduce the filter function in the gap equations, instead of introducing the cutoff energy used in previous studies. 
In Sec.~\ref{sec:bulkresults}, we show that the critical temperature obtained by the sparse modeling approach is quantitatively equivalent to that obtained by the standard BCS theory. 
In Sec.~\ref{sec:vortexresults}, we show that the sparse modeling approach can successfully reproduce a vortex-core shrinking, so-called Kramer-Pesch effect, reported in previous studies\cite{KramerPesch,HAYASHI201369,Hayashiimp}. 
In Sec.~\ref{sec:summary}, the summary is given.

\section{Quasiclassical theory of superconductivity} \label{sec:quasi}
\subsection{Definition}
We introduce the quasiclassical Green's function $\check{g}$ for a spin-singlet superconductor in equilibrium defined by 
\begin{align}
    \check{g}(i \omega_n,{\bm r},{\bm k}_{\rm F}) = \left( \begin{matrix}
    g & f \\
    - \bar{f} & -g
\end{matrix}
\right),
\end{align}
which is a function of the Matsubara frequency $\omega_n = \pi T (2n+1)$, the Fermi wave vector ${\bm k}_{\rm F}$, and the spatial coordinate ${\bm r}$. 
$\check{A}$ signifies the $2 \times 2$ matrix structure in the Nambu-Gor'kov particle-hole space. 
The definition of the quasiclassical Green's function is given by the contour integration: 
\begin{align}
    \check{g}(i \omega_n,{\bm r},{\bm k}_{\rm F}) &= -\oint \frac{d\xi_{\bm k}}{\pi i} \check{G}(i \omega_n,{\bm r},{\bm k}),
\end{align}
where $\check{G}(i \omega_n,{\bm r},{\bm k})$ is the Green's function in the Wigner representation. 
Here, the contour integral $\oint d\xi_{\bm k}$ means taking the contributions from poles close to the Fermi surface. 
The imaginary-time Green's function $\check{G}(\tau,{\bm r}_1,{\bm r}_2)$ is defined as 
\begin{align}
    \check{G}(\tau,{\bm r}_1,{\bm r}_2) &= 
    \left( 
    \begin{matrix}
    G(\tau,{\bm r}_1,{\bm r}_2) & F(\tau,{\bm r}_1,{\bm r}_2) \\
    F^{\dagger}(\tau,{\bm r}_1,{\bm r}_2) & \bar{G}(\tau,{\bm r}_1,{\bm r}_2)
    \end{matrix}
    \right),
\end{align}
where 
\begin{align}
 \check{G}(\tau,{\bm r}_1,{\bm r}_2) &\equiv -\langle T_{\tau} 
 \vec{\psi}(\tau,{\bm r}_1) \vec{\psi}^{\dagger}(0,{\bm r}_2)
\rangle.
\end{align}
Here, the operator is defined as $\vec{\psi}(\tau,{\bm r}) = (\psi_{\uparrow}(\tau,{\bm r}),\psi_{\downarrow}^{\dagger}(\tau,{\bm r}))^T$ and $\psi_{\alpha}(\tau,{\bm r})$ is the annihilation operator for the electron with spin $\alpha$ at ${\bm r}$.

The Eilenberger equation is the equation of motion for $ \check{g}(i \omega_n,{\bm r},{\bm k}_{\rm F})$, 
\begin{align}
    -i {\bm v}_{\rm F}({\bm k}_{\rm F}) \cdot {\bm \nabla} \check{g} = [i \tilde{\omega}_n \check{\tau}_3 - \check{\Delta}({\bm r},{\bm k}_{\rm F}),\check{g}], \label{eq:eilen}
\end{align}
supplemented by the normalization condition, 
\begin{align}
    \check{g}^2 = \check{1},
\end{align}
where $i \tilde{\omega}_n = i \omega_n + {\bm v}_{\rm F} \cdot \frac{e}{c} {\bm A}$, 
with ${\bm A}$ a vector potential and $\check{\tau}_3$ the Pauli matrix. 
$\check{\Delta}({\bm r},{\bm k}_{\rm F})$ is given by 
\begin{align}
   \check{\Delta}({\bm r},{\bm k}_{\rm F}) = \left( \begin{matrix}
   0 &  \Delta({\bm r},{\bm k}_{\rm F}) \\
   -  \Delta^{\ast}({\bm r},{\bm k}_{\rm F}) & 0
   \end{matrix} \right),
\end{align}
in the Nambu-Gor'kov space. 
For simplicity, we neglect the vector potential ${\bm A}$ in this paper. 

\subsection{Gap equation}
The order parameter for a spin-singlet superconductor is given by 
\begin{align}
    \Delta({\bm r}) &\equiv |g| \langle \psi_{\uparrow}({\bm r}) \psi_{\downarrow}({\bm r}) \rangle.
\end{align}
Here, we take into account that $g < 0$. 
The self-consistency equation for the order parameter becomes 
\begin{align}
     \Delta({\bm r}) =  -|g| F(\tau=0,{\bm r},{\bm r}) =   -|g| T \sum_{n=-\infty}^{\infty} F(i \omega_n,{\bm r},{\bm r}). \label{eq:gap}
\end{align}
With the use of the Wigner representation, the order parameter is given by 
\begin{align}
     \Delta({\bm r}) &=  -|g| T \sum_{n=-\infty}^{\infty} \int \frac{d^3 k}{(2 \pi)^3} F(i \omega_n,{\bm r},{\bm k}), \\
     &=\lambda \pi i T  \sum_{n=-\infty}^{\infty} \int \frac{d \Omega_{{\bm k}_{\rm F}}}{4 \pi} f(i \omega_n,{\bm r},{\bm k}_{\rm F}), \label{eq:gapmatsu} \\
     &=\lambda  \pi i  \int \frac{d \Omega_{{\bm k}_{\rm F}}}{4 \pi}   f(\tau = 0,{\bm r},{\bm k}_{\rm F}),
\end{align} 
where $\nu(0)$ is the density of states at the Fermi energy for normal states, $\lambda \equiv  |g| \nu(0)$, and we assume a spherical Fermi surface. 

\subsection{Homogeneous state}
Let us consider Green's functions for homogeneous state. 
The Gor'kov equation for a homogeneous case is given by 
\begin{align}
\left( \begin{matrix}
 i \omega_n - \xi_{\bm k} & -\Delta \\
-\Delta^{\ast} &  i \omega_n +  \xi_{\bm k}
\end{matrix}
\right) \check{G}(i \omega_n,{\bm k}) = \left( 
\begin{matrix}
1 & 0 \\
0 & 1
\end{matrix}
\right),
\end{align}
where $\xi_{\bm k}$ is the energy dispersion in normal state. 
The anomalous Green's function $F$ for homogeneous state is given by 
\begin{align}
    F(\omega_n) = -\frac{\Delta}{\xi_{\bm k}^2 + \omega_n^2 + |\Delta|^2}.
\end{align}
After the contour integration, the quasiclassical anomalous Green's function is given by 
\begin{align}
    f(\omega_n) &= \frac{\Delta}{i \sqrt{\omega_n^2 + |\Delta|^2}},
\end{align}
since the poles are located at $\xi_{\bm k} = \pm \sqrt{\omega_n^2 + |\Delta|^2}$. 

\section{Sparse modeling approach} \label{sec:spm}
\subsection{Lehmann representation}
We consider the Fermionic imaginary-time quantity, which is defined as the summation of the Fermion Matsubara frequencies:
\begin{align}
    I(\tau) = \frac{1}{\beta}\sum_{n=-\infty}^{\infty} I(i \omega_n) e^{- i \omega_n \tau},
\end{align}
The Lehmann representation of the quantity $I(\tau)$ is given by 
\begin{align}
     I(\tau) = -\int_{-\infty}^{\infty} d\omega K(\tau,\omega) \rho_{I}(\omega), \label{eq:lehmannrep}
\end{align}
where the spectral function $\rho_{I}(\omega)$ is defined as 
\begin{align}
    \rho_{I}(\omega) \equiv \lim_{\eta \rightarrow 0+ }\frac{-1}{2 \pi i } \left[ I(\omega + i \eta) - I(\omega - i \eta) \right].
\end{align}
Here, the kernel $K$ is defined as 
\begin{align}
     K(\tau,\omega) \equiv \frac{e^{- \tau \omega}}{1 + e^{- \beta \omega}}
\end{align}
for $0 < \tau < \beta$.

\subsection{Intermediate representation basis}
We introduce the cut-off $\omega_{\rm max}$ in the Lehmann representation given as 
\begin{align}
     I(\tau) \sim -\int_{-\omega_{\rm max}}^{\omega_{\rm max}} d\omega K(\tau,\omega) \rho_{I}(\omega). \label{eq:cut}
\end{align}
If the spectral function $\rho_{I}(\omega)$ is bounded in the interval $- \omega_{\rm max} < \omega < \omega_{\rm max}$, this expression becomes exact. 
For given $\omega_{\rm max}$ and $\beta$, the intermediate representation (IR) basis functions are defined through the singular value expansion: 
\begin{align}
     K(\tau,\omega) = \sum_{l=0}^{\infty} S_l U_l(\tau) V_l(\omega),
\end{align}
where 
\begin{align}
    \int_0^{\beta} d\tau U_l(\tau) U_{l'}(\tau) = \int_{-\omega_{\rm max}}^{\omega_{\rm max}}d\omega V_l(\omega) V_{l'}(\omega) = \delta_{ll'}.
\end{align}
Because the singular value $S_l$ decays exponentially with increasing $l$\cite{shinaokaPRB2017}, the function $I(\tau)$ can be expanded into a compact representation in terms of $N_{\rm IR}$ basis functions, such that in imaginary time and Matsubara frequencies, 
\begin{align}
    I(\tau) &\sim \sum_{l=0}^{N_{\rm IR}-1} I_l U_l(\tau), \label{eq:Itau-expans}\\
    I(i \omega_n) &\sim \sum_{l=0}^{N_{\rm IR}-1} I_l U_l(i \omega_n),\\
    U_l(i \omega_n) &= \int_0^{\beta} d\tau U_l(\tau) e^{i \omega_n \tau},
\end{align}
where $I_l$ are expansion coefficients, $U_l(\tau)$ is the $l$-th IR basis function. 
The expansion coefficients $I_l$ are related to the expansion coefficients of the spectral function as
\begin{align}
    I_l &= - S_l \int_{-\omega_\mathrm{max}}^{\omega_\mathrm{max}} d\omega \rho_I(\omega) V_l(\omega),
\end{align}
indicating that $|I_l|$ must decay as fast as $S_l$.

In practice, the coefficients $I_l$ can be computed by the fitting given as 
\begin{align}
    I_l = \mathop{\arg \min}_{I_l}  \sum_n \Bigl{|}I(i \omega_n) - \sum_{l=0}^{N_{\rm IR}-1} I_l U_{l}(\omega_n) \Bigl{|}^2.
\end{align}
With the use of the sparse sampling technique shown in Ref.~\cite{PhysRevB.101.035144},
the coefficients for the finite number of the functions $I(i \omega_n)$ can be computed by fitting data on carefully sampled $N_\mathrm{smpl}~(\gtrsim  N_{\rm IR})$ Matsubara frequencies.
The number of sampling points equals to the basis size or is slighly larger than the basis size by a few additional points.
One can easily generate the sampling points with the use of \texttt{SparseIR.jl} package\cite{sparseir}, since sampling Matsubara frequencies does depend on the kernel $K(\tau,\omega)$, not depend on function $I(i \omega_n)$. 

\section{Gap equation with the sparse modeling approach} \label{sec:gap}
\subsection{Without using the quasiclassical theory}
\begin{figure}[t]
(a)
  \begin{minipage}[b]{0.8 \columnwidth}
    \centering
    \includegraphics[keepaspectratio, width=\columnwidth]{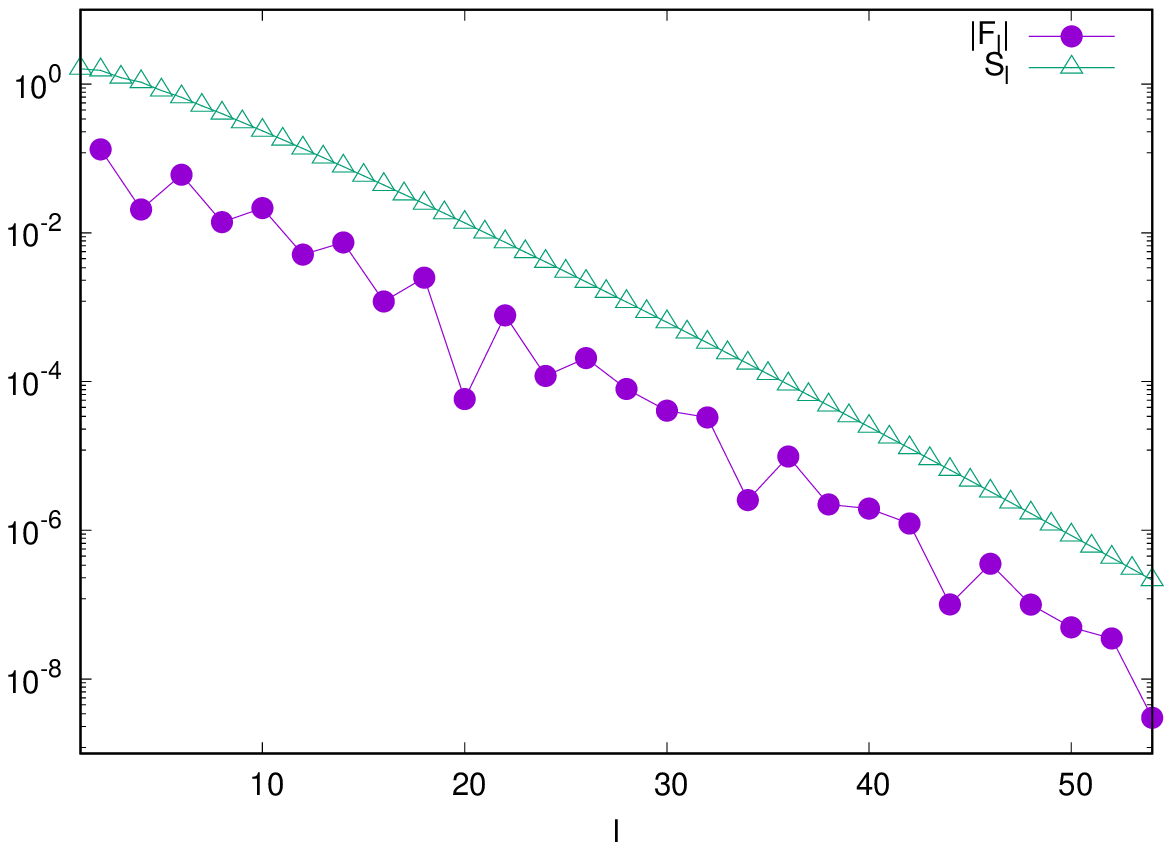}
  \end{minipage} \\
  (b)
  \begin{minipage}[b]{0.8 \columnwidth}
    \centering
    \includegraphics[keepaspectratio, width= \columnwidth]{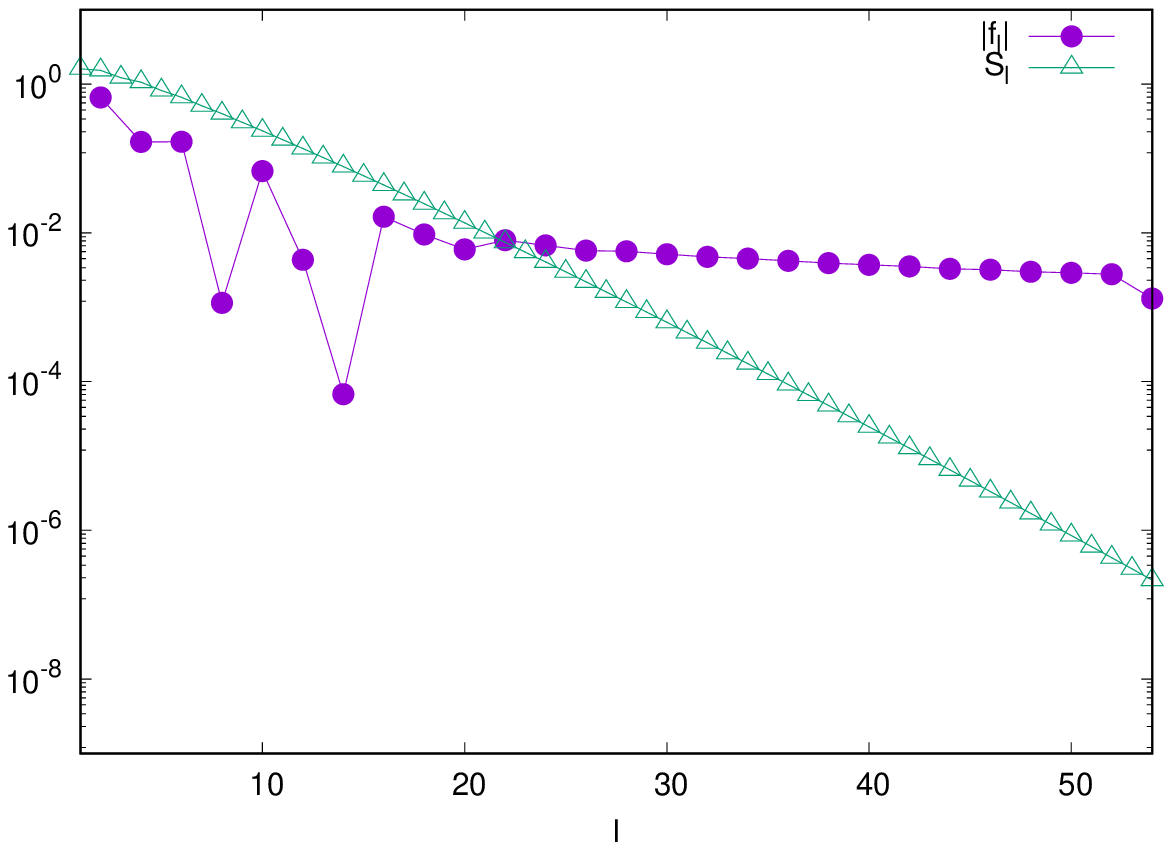}
  \end{minipage}
  \caption{IR coefficients for homogeneous state (a) without using the quasiclassical theory (b) with using the quasiclassical theory. $S_l$ are the singular values.}
  \label{fig:coeff}
\end{figure}

Let us consider the Lehmann representation of the gap equation.
Without using the quasiclassical theory, the gap equation (\ref{eq:gap}) is rewritten as 
\begin{align}
\Delta({\bm r}) =  |g|  \int_{-\infty}^{\infty} d\omega K(\tau=0,\omega) \rho_{F}(\omega,{\bm r},{\bm r}),
\end{align}
where 
\begin{align}
 \rho_{F}(\omega,{\bm r},{\bm r}) =\frac{1}{2 \pi i } \left[ F^{\rm R}(\omega,{\bm r},{\bm r}) - F^{\rm A}(\omega,{\bm r},{\bm r}) \right].
\end{align}
Here, $F^{\rm R({\bm A})}$ is a retarded (advanced) anomalous Green's function. 
With the use of the IR basis, the gap equation is given by 
\begin{align}
\Delta({\bm r}) =  |g|  \sum_{l=0}^{N_{\rm IR}-1} F_l U_l(\tau=0).
\end{align}
In homogeneous state, $\rho_{F}(\omega)$ is given by 
\begin{align}
     \rho_{F}(\omega) = \int \frac{d^3k}{(2 \pi)^3} \frac{ \delta(\omega -E_{\bm k} ) -  \delta(\omega +E_{\bm k}) }{E_{\bm k}},
\end{align}
where $E_{\bm k} \equiv \sqrt{\xi_{\bm k}^2 + |\Delta|^2}$. 
Introducing the cut-off $\omega_{\rm max}$ in Eq.~(\ref{eq:cut}) means removing quasiparticles with the energy $|E_{\bm k}| > \omega_{\rm max}$.  In inhomogeneous states, if the energy $\omega$ is much higher than the superconducting energy scale $|\Delta|$, the spectral function  $\rho_{F}(\omega,{\bm r},{\bm r})$ should have similar $\omega$ dependence. 
Therefore, we can use the IR basis to calculate the gap equations without using the quasiclassical theory. 
For example, we show the IR coefficients $F_l$ for homogeneous state with a momenta ${\bm k}$ in Fig.~\ref{fig:coeff}(a). Here, we consider $|\Delta|=1$, $\xi_{\bm k}=2$, and $T=0.01$. 
Figure \ref{fig:coeff}(a) shows that the gap equation in (\ref{eq:gap}) can be expanded by $50$ basis functions.

\subsection{With using the quasiclassical theory}
The gap equation with the use of the quasiclassical Green's function is given by 
\begin{align}
     \Delta({\bm r}) 
     &=  \lambda   \int \frac{d \Omega_{{\bm k}_{\rm F}}}{4 \pi} 
     \int_{-\infty}^{\infty} d\omega K(\tau=0,\omega) \rho_{f}(\omega,{\bm r},{\bm k}_{\rm F}), \label{eq:gapquasi}
\end{align} 
where 
\begin{align}
 \rho_{f}(\omega,{\bm r},{\bm k}_{\rm F}) =\frac{1}{2 } \left[ f^{\rm R}(\omega,{\bm r},{\bm k}_{\rm F}) - f^{\rm A}(\omega,{\bm r},{\bm k}_{\rm F}) \right].
\end{align}
Here, $f^{\rm R({\bm A})}$ is a retarded (advanced) quasiclassical anomalous Green's function. 

In homogeneous state, the spectral function $ \rho_{f}(\omega,{\bm k}_{\rm F})$ is given by (See, Appendix \ref{app:lehmann}.)
\begin{align}
     \rho_{f}(\omega,{\bm k}_{\rm F}) &= \frac{\Delta}{\sqrt{\omega^2 - |\Delta|^2}} \Theta(\epsilon^2 - |\Delta|^2).
\end{align}
We should note that this spectral function is not bounded in finite intervals. 
Therefore, the integration  in Eq.~(\ref{eq:gapquasi}) depends on the cut-off frequency $\omega_{\rm max}$. 
In this case, the IR basis is not valid for reproducing the Matsubara Green's function: The IR coefficients $f_l$ do not decay with increasing the index $l$, as shown in Fig.~\ref{fig:coeff}(b).
Here, we consider $|\Delta|=1$, $\omega_{\rm cut} = 100$ and $T=0.01$. 

The reason of the unconvergence originates from the fact that both the Matsubara summation in Eq.~(\ref{eq:gapmatsu}) and the real-frequency integration in Eq.~(\ref{eq:gapquasi}) diverge. 
In the conventional quasiclassical theory, to avoid this divergence,  the cut-off for the Matsubara summation is introduced\cite{kopnin2001theory}: 
\begin{align}
     \Delta({\bm r}) &\sim \lambda \pi i T  \sum_{-\omega_{\rm c} < \omega_n < \omega_{\rm c}} \int \frac{d \Omega_{{\bm k}_{\rm F}}}{4 \pi} f(i \omega_n,{\bm r},{\bm k}_{\rm F}). \label{eq:gapcut}
\end{align} 
Here, the cut-off energy $\omega_{\rm c}$ is usually set to $10-100 \Delta_0$\cite{Hayashiimp}, where $\Delta_0$ is a superconducting order parameter at zero temperature. 
However, we point out that the effect of this cut-off in the Matsubara summation is not equivalent to that of the cut-off in the real-frequency integration, because the Matsubara summation and real-frequency integration are connected by the residue theorem (see, Appendix \ref{app:lehmann}.). 
Therefore, Eq.~(\ref{eq:gapcut}) can be regarded as the equation with the filter function $\Theta( \omega_{\rm c}^2- \omega_n^2)$.
Note that introducing the hard cut-off in the Matsubara summation does not have a reasonable physical interpretation.

\begin{figure}[t]
    \centering
    \includegraphics[keepaspectratio, width=\columnwidth]{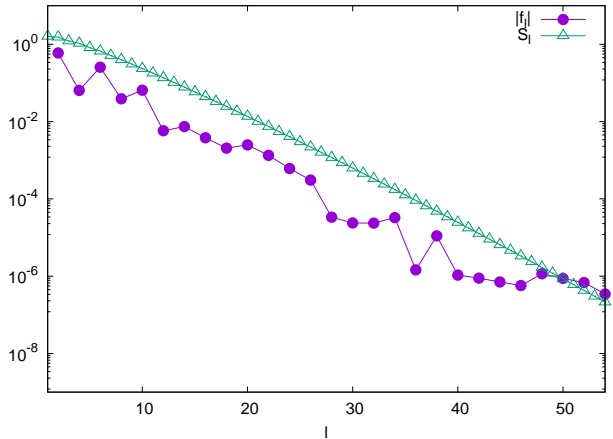}
  \caption{IR coefficients for homogeneous state with using the quasiclassical theory with the filter function $V(\omega_n)$. $S_l$ are the singular values.  }
  \label{fig:coeff_f}
\end{figure}

To avoid the divergence in the Matsubara summation, we can introduce a different filter function $V(\omega_n)$: 
\begin{align}
     \Delta({\bm r}) 
     &=\lambda \pi i T  \sum_{n=-\infty}^{\infty} \int \frac{d \Omega_{{\bm k}_{\rm F}}}{4 \pi} V(\omega_n) f(i \omega_n,{\bm r},{\bm k}_{\rm F}), \label{eq:gapfiltered} 
\end{align} 
In this paper, we introduce a filter function $V(\omega_n) = \omega_{\rm D}^2/(\omega_n^2 + \omega_{\rm D}^2)$. 
In homogeneous state, the gap equation with this filter in the Lehmann representation is given by (see, Appendix \ref{app:filter}), 
\begin{align}
     \Delta
     &=\lambda \int \frac{d \Omega_{{\bm k}_{\rm F}}}{4 \pi} 
     \int_{-\infty}^{\infty} d\omega K(\tau=0,\omega) \rho_{f,{\rm filter}}(\omega,{\bm k}_{\rm F}), \label{eq:gapquasiV}
\end{align} 
where
\begin{align}
 \rho_{f,{\rm filter}}(\omega,{\bm k}_{\rm F}) = {\cal P} \frac{\omega_{\rm D}^2}{\omega_{\rm D}^2 - \omega^2}  \rho_{f}(\omega,{\bm k}_{\rm F}). \label{eq:gapv}
\end{align}
Here, ${\cal P}$ denotes the Cauchy principal value where the points $\omega = \pm \omega_{\rm D}$ are avoided in the real-frequency integration (see, appendix). 
If $\omega_{\rm D} \gg |\Delta|$, the filter function becomes $V(\omega) \sim 1$ in the low energy region. 
In the high-energy region $(|\omega| > \omega_D)$, the filter function can be regarded as the effective repulsive interaction, which is physically reasonable since the attractive interaction exists only in the low energy region. 
Thus, we can introduce the IR basis with the cut-off $\omega_{\rm max}$ that is much higher than $\omega_{\rm D}$. 
As shown in Fig.~ \ref{fig:coeff_f}, the IR coefficients $f_l$ decay exponentially with increasing the index $l$. 
Here, we consider $T = 0.01$, $\Delta = 1$, $\omega_{\rm D} = 10$, and $\omega_{\rm max} = 100$.

\section{Numerical demonstrations for bulk state} \label{sec:bulkresults}
We show the temperature dependence of the superconducting order parameter of $s$-wave superconductor in bulk state.
To normalize the order parameter, we determine the pairing interaction $\lambda$ that gives $\Delta = 1$ at $T = 0.01$. We regard $\Delta(T = 0.01)$ as the zero-temperature order parameter $\Delta_0$. 
In the conventional quasiclassical theory of superconductivity, the pairing interaction is calculated by 
\begin{align}
    \frac{1}{\lambda_{\rm conv}} = \sum_{- \omega_{\rm c} < \omega_n < \omega_{\rm c}} \frac{\pi}{\sqrt{\omega_n^2 + 1}}.
\end{align}
According to the previous study\cite{Hayashiimp,HAYASHI201369}, we consider $\omega_{\rm c} = 10\Delta_0$. For a comparison, we also consider the case with $\omega_{\rm c} = 100\Delta_0$.
In our quasiclassical theory with the sparse modeling, the pairing interaction is calculated by  \begin{align}
    \frac{1}{\lambda_{\rm SpM}} = \sum_{n = -\infty}^{\infty} \frac{\omega_{\rm D}^2}{\omega_n^2 + \omega_{\rm D}^2}\frac{\pi }{\sqrt{\omega_n^2 + 1}}.
\end{align}
Here, we consider $\omega_{\rm D} = 10 \Delta_0$ and $\omega_{\rm D} = 100 \Delta_0$. We set the cut-off energy $\omega_{\rm max} = 10\omega_{\rm D}$. 
The IR basis and sampling Matsubara frequencies are calculated by \texttt{SparseIR.jl} package written in the Julia language. 

\begin{figure}[t]
(a)
  \begin{minipage}[b]{0.8 \columnwidth}
    \centering
    \includegraphics[keepaspectratio, width=0.98 \columnwidth]{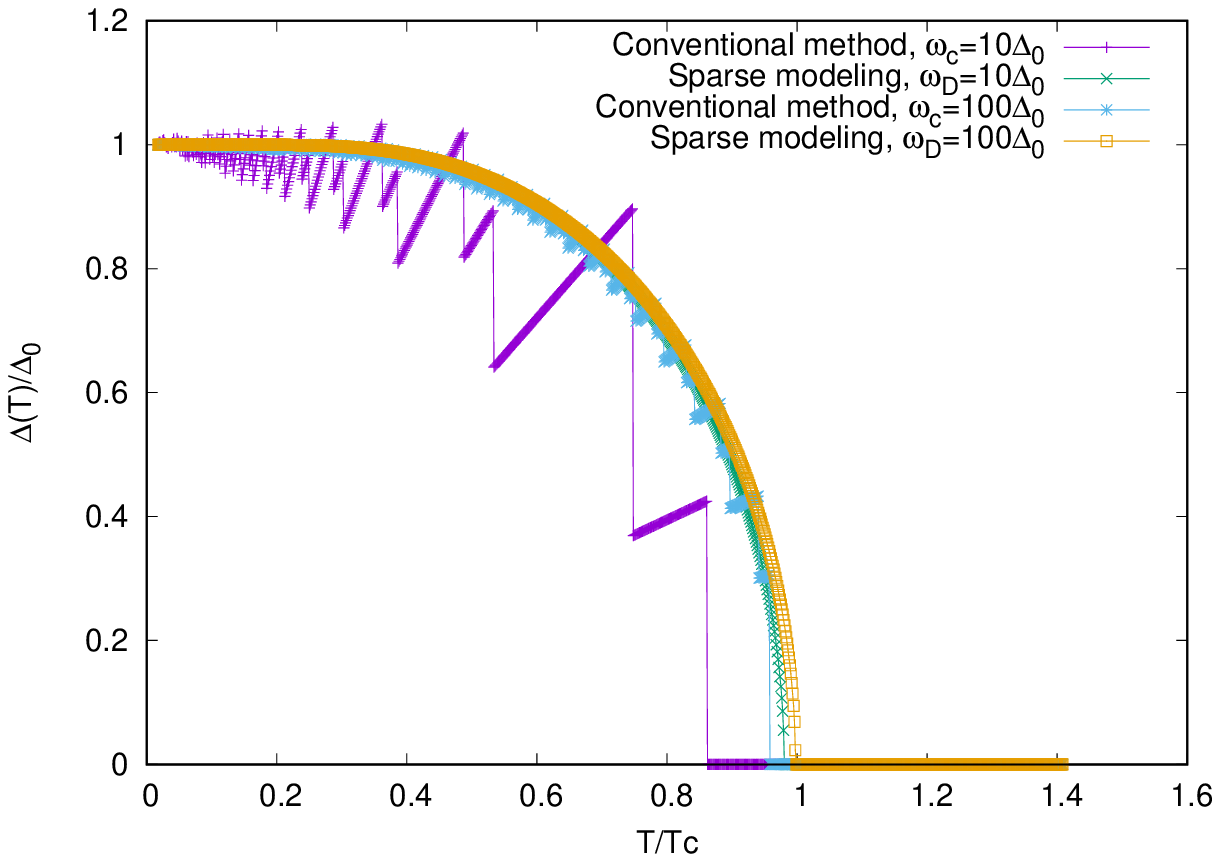}
  \end{minipage} \\
  (b)
  \begin{minipage}[b]{0.8 \columnwidth}
    \centering
    \includegraphics[keepaspectratio, width= \columnwidth]{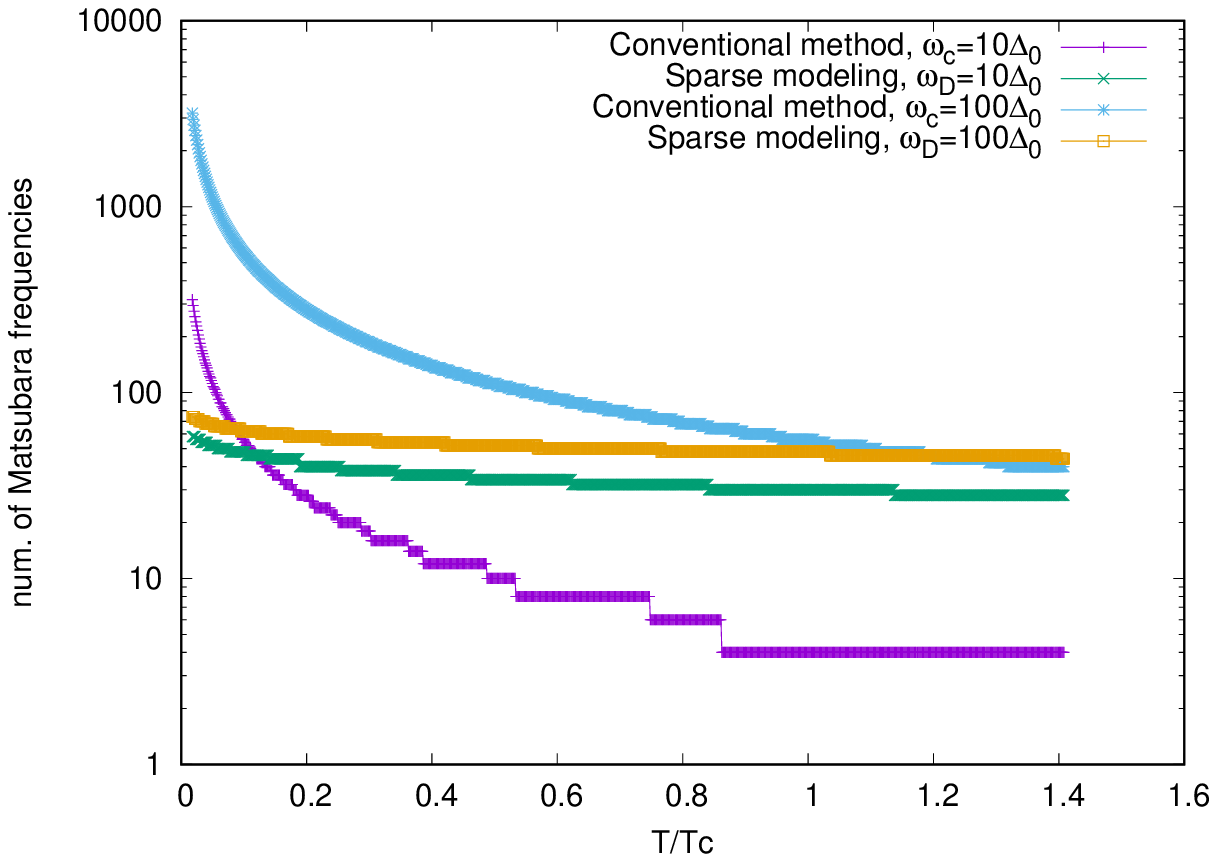}
  \end{minipage}
  \caption{(a) Temperature dependence of the superconducting order parameter in bulk state. We set $T_{\rm c} = \Delta_0/1.76$. (b) Temperature dependence of the number of Matsubara frequencies. }
  \label{fig:gaptemp}
\end{figure}

Figure \ref{fig:gaptemp}(a) shows that our quasiclassical theory with the sparse modeling successfully reproduces the critical temperature  $T_{\rm c} = \Delta_0/1.76$, derived by the standard BCS theory. 
As shown in Appendix \ref{app:integral}, if the cut-off energy $\omega_{\rm D}$ is large enough ($\omega_{\rm D} \gg |\Delta|$), the integral in the gap equation with the Lehmann representation (\ref{eq:gapv}) becomes the standard result of the BCS theory.  
On the other hand, the temperature dependence of the order parameter calculated by the conventional quasiclassical theory has non-monotonic behavior. 
This strange behavior in Fig.~\ref{fig:gaptemp} originates from introducing the cut-off energy in the Matsubara summation. 
In conventional quasiclassical theory of superconductivity,  the cut-off number of the Matsubara frequencies $n_{\rm c}$, determined by $\omega_{\rm c} = \pi T (2 n_{\rm c} + 1)$,  decreases discretely with increasing temperature.   
For example, at $T = 0.8 T_{\rm c} = (0.8/1.76) \Delta_0$, there are only six Matsubara frequencies less than $\omega_{\rm c} = 10\Delta_0$.
The Matsubara summation with six Matsubara frequencies can not reproduce the real-frequency integration.

\section{Numerical demonstration for vortex state} \label{sec:vortexresults}
We consider a two-dimensional $s$-wave superconductor with a vortex. 
As shown in Fig.~\ref{fig:gaptemp}(b), in low temperature region, the number of the Matsubara frequencies that we have to consider in our quasiclassical theory with the sparse modeling is much smaller than that in the conventional quasiclassical theory. 
Since the calculation cost of the self-consistent calculation is proportional to the number of the Matsubara frequencies, our method is much faster than the conventional method. 
In high temperature region, the temperature dependence of the order parameter in our theory is more natural than that in the conventional quasiclassical theory as shown in Fig.~\ref{fig:gaptemp}(a). 
Therefore, the our sparse modeling approach is appropriate for studying the temperature dependence of superconductors. 

Kramer and Pesch\cite{KramerPesch} have pointed out that the radius of a vortex core decreases proportionally to the temperature in low temperatures, much stronger than anticipated from the temperature dependence of the coherence length. 
The strong shrinking of the vortex core, the so-called Kramer-Pesch effect, has been studied in various kinds of superconductors\cite{Hayashiimp,HAYASHI201369,doi:10.1143/JPSJ.70.3368}. 
In this section, we reproduce the Kramer-Pesch effect with the use of the sparse modeling approach. 

\subsection{Method}
The Eilenberger equation (\ref{eq:eilen}) can be solved by the Riccati parametrization\cite{Schopohl,PhysRevB.61.9061,nagai2006,PhysRevB.86.094526}. 
The quasiclassical Green's function is expressed as 
\begin{align}
    \check{g}(i \omega_n,{\bm r},{\bm k}_{\rm F}) = \frac{{\rm sgn}(\omega_n)}{1 + a b} \left( \begin{matrix}
        1-ab & i2a\\
        -i2b & -(1-ab)
    \end{matrix} \right).
\end{align}
The variables $a(i \omega_n,{\bm r},{\bm k}_{\rm F})$ and $b(i \omega_n,{\bm r},{\bm k}_{\rm F})$ are independently determined by solving the Riccati equations,
\begin{align}
    {\bm v}_{\rm F} \cdot {\bm \nabla} a &= - 2 \omega_n a - \Delta^{\ast} a^2 + \Delta, \\
      {\bm v}_{\rm F} \cdot {\bm \nabla} b &= 2 \omega_n a + \Delta a^2 - \Delta^{\ast}.
\end{align}
These differential equations are solved along a straight line parallel to ${\bm v}_{\rm F}$ by using the bulk solutions as initial values. 
In the case of $\omega_n > 0$, the initial values are given as 
\begin{align}
    a_{-\infty} &= \frac{- \omega_n + \sqrt{\omega_n^2 + |\Delta|^2}}{\Delta^{\ast}}, \label{eq:ap} \\
    b_{-\infty} &= \frac{- \omega_n + \sqrt{\omega_n^2 + |\Delta|^2}}{\Delta}. \label{eq:bp}
\end{align}
In the case of $\omega_n < 0$, the initial values are given as 
\begin{align}
    a_{+\infty} &= \frac{-1}{b_{+\infty}} =  \frac{- \omega_n - \sqrt{\omega_n^2 + |\Delta|^2}}{\Delta^{\ast}},  \label{eq:am}\\
    b_{+\infty} &= \frac{-1}{a_{-\infty}} = \frac{- \omega_n - \sqrt{\omega_n^2 + |\Delta|^2}}{\Delta}. \label{eq:bm}
\end{align}
A stable numerical solution for the valuable $a$ ($b$) is obtained by solving the Riccati equation in forward (backward) direction along the straight line for $\omega_n > 0$\cite{PhysRevB.86.094526}. 
For $\omega_n <0$, the equation for $a$ ($b$) is solved in backward (forward) direction. 
In this paper, these Riccati equations are solved by \texttt{DifferentialEquations.jl}, a package written in the Julia language. 
We use the Vern8 algorithm, Verner's ``Most Efficient'' 8/7 Runge-Kutta method. 
The other technical details for solving the Riccati equations are shown in Ref.~\cite{HAYASHI201369}. 

To normalize the order parameter far from a vortex core $\Delta(T,{\bm r} > {\bm r}_c) = \Delta(T,{\bm r} \rightarrow \infty) = 1$, we change the pairing interaction with changing temperature. 
Here, we set ${\bm r}_c = 10$ and the unit of the length scale in the Riccati equations is $\xi_0(T) \equiv |{\bm v}_{\rm F}|/\Delta(T,{\bm r} \rightarrow \infty)$.
We consider a circular Fermi surface ($|{\bm v}_{\rm F}|=1$) for simplicity.
The pairing interaction is given by 
\begin{align}
    \lambda_{\rm conv}(T) &= \left[ \sum_{- \omega_{\rm c} < \omega_n(T) < \omega_{\rm c}} \frac{\pi}{\sqrt{\omega_n(T)^2 + 1}} \right]^{-1}, \\
    \lambda_{\rm SpM}(T) &=\left[ \sum_{n = -\infty}^{\infty} \frac{\omega_{\rm D}^2}{\omega_n(T)^2 + \omega_{\rm D}^2}\frac{\pi }{\sqrt{\omega_n(T)^2 + 1}} \right]^{-1}.
\end{align}
In the high Matsubara frequency region ($|\omega_n| > 100$), we do not solve the Riccati equations numerically and use the bulk solutions (\ref{eq:ap})-(\ref{eq:bm}). 

\begin{figure}[t]
    \centering
    \includegraphics[keepaspectratio, width=\columnwidth]{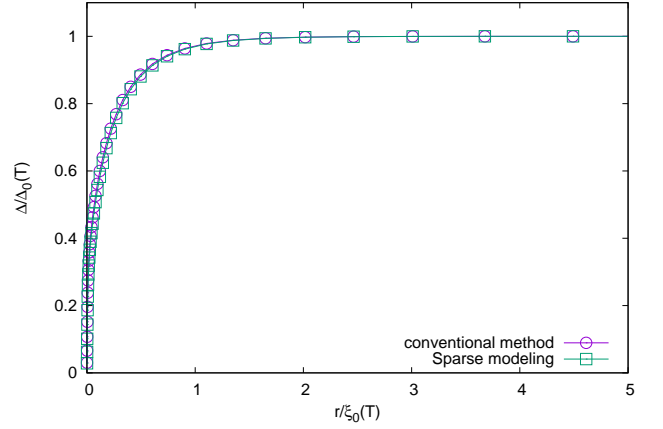}
  \caption{Radial dependence of the superconducting order parameter around a vortex at $T = 0.02T_{\rm c}$ with $\omega_c = \omega_{\rm D} = 10\Delta_0$, and $\omega_{\rm max} = 10 \omega_{\rm D}$. }
  \label{fig:comparison}
\end{figure}

\begin{figure}[t]
    \centering
    \includegraphics[keepaspectratio, width=\columnwidth]{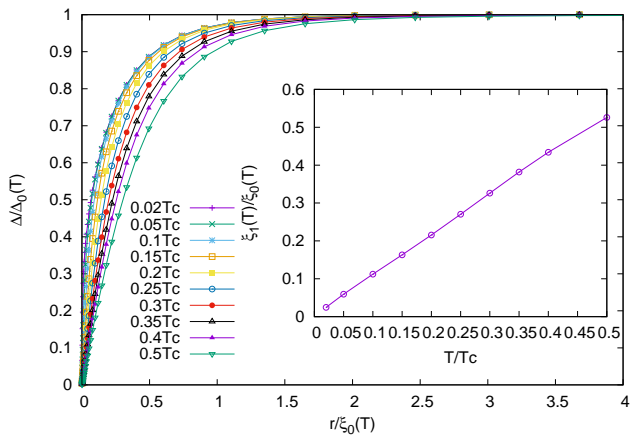}
  \caption{Temperature dependence of the radial distribution of the superconducting order parameter around a vortex. (Inset) Temperature dependence of the vortex core radius. We use the sparse modeling approach with  $\omega_{\rm D} = 10\Delta_0$, and $\omega_{\rm max} = 10 \omega_{\rm D}$. }
  \label{fig:kp}
\end{figure}

\subsection{Results}
The converged order parameter distribution around a vortex is shown in Fig.~\ref{fig:comparison}. 
The radial dependence of the order parameter calculated by the sparse modeling approach is quantitatively equivalent to that calculated by the conventional method. 
The result shows that the filter function $V(\omega_n)$ is appropriate in inhomogeneous superconductors. 
The number of the Matsubara frequencies that we have to consider in the sparse modeling approach is $58$ with the parameters $\omega_{\rm D}=10\Delta_0$ and $\omega_{\rm max} = 10\omega_{\rm D}$, while the number in the conventional method is 280.
Since the number of the Matsubara frequencies is proportional to elapsed time, the elapsed time in the sparse modeling approach is 4 times or more shorter than that in the conventional method. 
If one considers $\omega_{\rm c} = \omega_{\rm D} = 100\Delta_0$, the ratio of the number of Matsubara frequency is $74/2816$. 

Figure \ref{fig:kp} shows the temperature dependence of the radial distribution of the superconducting order parameter around a vortex.
One can clearly see the Kramer-Pesch effect, the shrinkage of the vortex core with decreasing temperature.
Here, the vortex core radius $\xi_1$ determined as\cite{HAYASHI201369}, 
\begin{align}
    \frac{1}{\xi_1} = \frac{1}{\Delta(r \rightarrow \infty)} \lim_{r \rightarrow 0} \frac{\Delta(r)}{r}.
\end{align}
The inset in Fig.~\ref{fig:kp} shows that $\xi_1$ is proportional to temperature. 
This result obtained by the sparse modeling approach is quantitatively equivalent to the result reported in Ref.~\cite{Hayashiimp}. 

The big advantage of the sparse modeling approach is the weak temperature dependence of the number of the Matsubara frequencies. 
In the conventional method, the number of the Matsubara frequency is proportional to $1/T$. 
On the other hand, the sparse modeling approach, the number is proportional to $-\log T$\cite{PhysRevB.98.035104}. 
Therefore, the sparse modeling approach is appropriate in all theoretical methods based on the Matsubara formalism in the quasiclassical theory of superconductivity. 

\section{Summary} \label{sec:summary}
We proposed the sparse modeling approach for quasiclassical theory of superconductivity, for self-consistently solving the gap equations. 
We introduced the filter function $V(\omega_n)$ to use the IR basis in the gap equation. 
We solved the gap equations in bulk and vortex states and showed that all result calculated by the sparse modeling approach is quantitatively equivalent to that obtained by the conventional quasiclassical theory. 
The number of Matsubara frequencies that we have to consider in the gap equations is drastically reduced in the lower temperature region. 
We claimed that the sparse modeling approach is appropriate in all theoretical methods based on the Matsubara formalism in the quasiclassical theory of superconductivity.

\begin{acknowledgments}
YN was partially supported by JSPS- KAKENHI Grant Numbers 20H05278, 22H04602 and 22K03539. The calculations were partially performed by the supercomputing system HPE SGI8600 at the Japan Atomic Energy Agency.
The code with \texttt{SparseIR.jl} and \texttt{DifferentialEquations.jl} is written in the Julia language 1.7.2. 
HS was supported by JSPS KAKENHI Grant Numbers 21H01003 and 21H01041 as well as PRESTO Grant No. JPMJPR2012.
\end{acknowledgments}

\appendix

\section{Lehmann representation} \label{app:lehmann}
We derive the Lehmann representation Eq.~(\ref{eq:lehmannrep}).
We introduce a Fermionic imaginary-time quantity defined as 
\begin{align}
    I(\tau) = \frac{1}{\beta}\sum_{n=-\infty}^{\infty} I(i \omega_n) e^{- i \omega_n \tau},
\end{align}
where $\omega_n \equiv \pi T (2n + 1)$ is the Fermionic Matsubara frequency. 
We also introduce the complex function defined as $J(z) \equiv I(z)/(1+e^{\beta z})$.
Since the function $J(z)$ has the poles $z = i \omega_n$ ($n = 0,\pm 1,\pm 2,\cdots$), the function $I(\tau)$ is given by 
\begin{align}
     I(\tau)  = \frac{1}{2 \pi i} \oint_{C_+^{(i)}} dz \frac{e^{-z \tau}}{1+e^{\beta z} } I(z) + \frac{1}{2 \pi i} \oint_{C_-^{(i)}} dz \frac{e^{-z \tau}}{1+e^{\beta z} } I(z),
\end{align}
Here, the contour $C_+^{(i)}$ encircles all the poles of $1/(1+e^{\beta z})$ in the upper complex half plane of $z$ while the contour $C_-^{(i)}$ encircles all the poles of $1/(1+e^{\beta z})$ in the lower complex half plane of $z$. 
If the function $I(z)$ is an analytical function in both uppter and lower half plane, we can shift the contours in such a way that  $C_+^{(i)}$ transforms into  $C_+^{(r)}$ and goes along the real axis of $z$ from $-\infty$ to $+\infty$ and  $C_-^{(i)}$ transforms into  $C_-^{(r)}$ which goes along the real axis of $z$ from $+\infty$ to $-\infty$. 
Therefore, the  Lehmann representation for the function $I(\tau)$ is given by 
\begin{align}
     I(\tau) = -\int_{-\infty}^{\infty} d\omega K(\tau,\omega) \rho_{I}(\omega),
\end{align}
where 
\begin{align}
    \rho_{I}(\omega) \equiv \lim_{\eta \rightarrow 0+ }\frac{-1}{2 \pi i } \left[ I(\omega + i \eta) - I(\omega - i \eta) \right].
\end{align}

\section{Analytic continuation} \label{app:ac}
We derive the retarded and advanced quasiclassical anomalous Green's function in bulk with the use of the analytic continuation. 
The quasiclassical anomalous Matsubara Green's function in bulk is given by 
\begin{align}
    f(i \omega_n) = \frac{\Delta}{i \sqrt{\omega_n^2 + |\Delta|^2}}.
\end{align}
By replacing $i \omega_n$ with a complex variable $z$, the Green's function on the complex plane is defined as 
\begin{align}
    f(z) = \frac{\Delta}{i \sqrt{-z^2 + |\Delta|^2}}=  \frac{\Delta}{i \sqrt{-(z - |\Delta|)(z + |\Delta|)}}.
\end{align}
To obtain the retarded and advanced Green's function, we have to define the sign of the square root in the above equation. 
By defining the following quantities: 
\begin{align}
    z + |\Delta| &= |z + |\Delta|| e^{i \theta_+}, \\ 
    z - |\Delta| &= |z - |\Delta|| e^{i \theta_-},
\end{align}
the square root becomes 
\begin{align}
   \sqrt{-(z - |\Delta|)(z + |\Delta|)}&=   \left( e^{\pm i \pi} |z + |\Delta|| e^{i \theta_+}  |z - |\Delta|| e^{i \theta_-} \right)^{\frac{1}{2}},\\
   &=  |z + |\Delta||^{\frac{1}{2}}  |z - |\Delta||^{\frac{1}{2}}  e^{i \frac{\theta_+ + \theta_- \pm \pi}{2}}.  
\end{align}
Here, we fix the phase $(\theta_+ + \theta_- \pm \pi)/2 \rightarrow (\theta_+ + \theta_- - \pi)/2$ to reproduce the Green's function with Matsubara frequency on the upper half plane, since the phases $\theta_+$ and $\theta_-$ becomes $\theta_+ = \theta_- = \pi/2$ in the limit of the high Matsubara frequency on the upper half plane ($z \rightarrow i \infty$)). 
Thus, the function $f(z)$ is given by 
\begin{align}
      f(z) = \frac{\Delta}{i |z + |\Delta||^{1/2} |z - |\Delta||^{1/2} } e^{-i (\theta_+ + \theta_- - \pi)/2}.
\end{align}
The retarded (advanced) Green's function is obtained by replacing $z$ with $\omega + i \eta$ ($\omega - i \eta$). 
In the case of $\omega > |\Delta|$, the phases are $\theta_+ = \theta_- = 0$  for the retarded Green's function and  $\theta_+ = \theta_- = 2\pi$ for the advanced Green's function. 
In the case of $0 < \omega < |\Delta|$, the phases are  $\theta_+ = 0$ and $\theta_- = \pi$ for both retarded and advanced Green's functions. 
Thus, the retarded and advanced Green's functions are given as 
\begin{align}
     f^R(\omega) &= 
     \begin{cases}
\frac{\Delta}{i \sqrt{-\omega^2 + |\Delta|^2}}, & \omega > |\Delta|, \\
- \frac{\Delta}{i \sqrt{ |\Delta|^2-\omega^2 }}, & 0< \omega < |\Delta|,
\end{cases}\\
f^A(\omega) &= 
     \begin{cases}
- \frac{\Delta}{i \sqrt{-\omega^2 + |\Delta|^2}}, & \omega > |\Delta|, \\
- \frac{\Delta}{i \sqrt{ |\Delta|^2-\omega^2 }}, & 0< \omega < |\Delta|.
\end{cases}
\end{align}
With the use of the similar discussion for $\omega < 0$, we obtain 
\begin{align}
    f^R(\omega) &= - f^A(\omega), \: \: |\omega| > |\Delta|, \label{eq:a1} \\
    f^R(\omega) &=  f^A(\omega), \: \: |\omega| < |\Delta|. \label{eq:a2}
\end{align}

\section{Lehmann representation with the filter function} \label{app:filter}
We derive Eqs.~(\ref{eq:gapquasiV}) and (\ref{eq:gapv}). 
In homogeneous states, the Lehmann representation is given as  
\begin{align}
     \Delta
     &=\frac{\lambda}{2} \int \frac{d \Omega_{{\bm k}_{\rm F}}}{4 \pi} 
     \int_{-\infty}^{\infty} d\omega K(\tau,\omega)f^{\rm R}(\omega)
     \left(V^{\rm R}(\omega)+ V^{\rm A}(\omega) \right) .
\end{align}
Here, we use Eqs.~(\ref{eq:a1}) and (\ref{eq:a2}). 
The sum of the retarded and advanced filter function is expressed as 
\begin{align}
    V^{\rm R}(\omega)  + V^{\rm A}(\omega) &=   \lim_{\eta \rightarrow 0+} V(\omega + i \eta) +  V(\omega - i \eta), \\
    &=  2 {\cal P} \frac{\omega_c^2}{\omega_c^2- \omega^2}.
\end{align}
Thus, we obtain Eqs.~(\ref{eq:gapquasiV}) and (\ref{eq:gapv}).

\section{Validity of the filter function} \label{app:integral}
We discuss the validity of the filter function $V(\omega_n)$. 
In homogeneous system at zero temperature, the gap equation in the conventional BCS theory is given as 
\begin{align}
    \frac{1}{ \lambda_{\rm conv}} = \int_{-\Omega}^{-1} d\omega \frac{1}{\sqrt{\omega^2 - 1}} = \int_{1}^{\Omega} d\omega \frac{1}{\sqrt{\omega^2 - 1}}, \label{eq:appgap}
\end{align}
where $\Omega$ is the cut-off energy and we consider $\Delta(T=0) = 1$ as a unit. With the use of the filtering function $V(\omega_n) = \omega_{\rm D}^2/(\omega_{\rm D}^2 + \omega_n^2)$, the gap equation is given as  
\begin{align}
    \frac{1}{ \lambda_{\rm SpM}} =  {\cal P} \int_{1}^{\infty} d\omega \frac{\omega_{\rm D}^2}{\omega_{\rm D}^2 - \omega^2} \frac{1}{\sqrt{\omega^2 - 1}}. \label{eq:appgapspm}
\end{align}
We compare Eq.~(\ref{eq:appgapspm}) with Eq.~(\ref{eq:appgap}). 
We introduce the energy $\omega_0$ which satisfies $1 \ll \omega_0 \ll \omega_{\rm D}$. 
With the use of this energy, we have 
\begin{align}
     \frac{1}{ \lambda_{\rm SpM}} =   \int_{1}^{\omega_0} d\omega  \frac{1}{\sqrt{\omega^2 - 1}} +  {\cal P} \int_{\omega_0}^{\infty} d\omega \frac{\omega_{\rm D}^2}{\omega_{\rm D}^2 - \omega^2} \frac{1}{\omega}.
\end{align}
The Cauchy principal value is given as  
\begin{align}
     {\cal P} \int_{\omega_0}^{\infty} d\omega \frac{\omega_{\rm D}^2}{\omega_{\rm D}^2 - \omega^2} \frac{1}{\omega} &=  \lim_{\epsilon \rightarrow 0+} \left[ 
     \int_{\omega_0}^{\omega_{\rm D}-\epsilon} d\omega \frac{\omega_{\rm D}^2}{\omega_{\rm D}^2 - \omega^2} \frac{1}{\omega} \right. \nonumber \\
     &- \left.  \int_{\omega_{\rm D}+\epsilon }^{\infty} d\omega \frac{\omega_{\rm D}^2}{ \omega^2 - \omega_{\rm D}^2 } \frac{1}{\omega} \right], \\
      &= \lim_{\epsilon \rightarrow 0+} \left[ \log \frac{\omega}{\sqrt{\omega_{\rm D}^2 - \omega^2}} \right]_{\omega_0}^{\omega_{\rm D}-\epsilon} \nonumber \\
      &+  \lim_{\epsilon \rightarrow 0+} \left[  \log \frac{\omega}{\sqrt{ \omega^2- \omega_{\rm D}^2 }} \right]_{\omega_{\rm D}+\epsilon }^{\infty}, \\
      &= -\log \frac{\omega_0}{\sqrt{\omega_{\rm D}^2 - \omega_0^2}},\\
      &\sim \log \frac{\omega_{\rm D}}{\omega_0}.
\end{align}
On the other hand, we have 
\begin{align}
     \frac{1}{ \lambda_{\rm conv}} &= \int_{1}^{\omega_0} d\omega \frac{1}{\sqrt{\omega^2 - 1}} +  \int_{\omega_0}^{\Omega} d\omega \frac{1}{\omega}, \\
     &= \int_{1}^{\omega_0} d\omega \frac{1}{\sqrt{\omega^2 - 1}} + \log \frac{\Omega}{\omega_0}.
\end{align}
Therefore, if we set $\omega_{\rm D} = \Omega$, $1/\lambda_{\rm SpM}$ becomes $1/\lambda_{\rm conv}$. 
In finite temperature, by introducing the energy $\omega_0 \gg T$, we can also show that the integral values in both gap equations becomes same. 
This means that the critical temperature obtained by both gap equations is same when the cut-off energy $\omega_{\rm D}$ is large enough ($\omega_{\rm D} \gg \Delta_0$).

\bibliography{quasiir}

\end{document}